\documentclass[letter]{aa} 

\usepackage{graphicx}
\usepackage{txfonts}
\newcommand{\Ms}{$\textrm{M}_{\odot}$}
\newcommand{\SFR}{\textrm{M}$_{\odot}$~\textrm{yr}$^{-1}$}
\newcommand{\kms}{$\textrm{km~s$^{-1}$}$}
\newcommand{\Ha}{H$\alpha$} 

\begin{document}

   \title{Star formation in outer rings of S0 galaxies. III}

   \subtitle{UGC 5936 -- an S0 with currently accreted satellite matter.}

   \author{O. Sil'chenko
          \inst{1}
          \and
          A. Moiseev
          \inst{2,1}
          }

   \institute{Sternberg Astronomical Institute of the Lomonosov Moscow
             State University, University av. 13, 119234 Russia\\
             \email{olga@sai.msu.su}
          \and
             Special Astrophysical Observatory
             of the Russian Academy of Sciences,
             Nizhnij Arkhyz, 369167 Russia\\
              \email{moisav@gmail.com}
             }

   \date{Received  .., 2020; accepted .., 2020}

 
  \abstract
   {}
   {Though S0 galaxies are usually thought to be `red and dead', they
   demonstrate often star formation organised in ring structures. We try
   to clarify the nature of this phenomenon and its difference from star
   formation in spiral galaxies. The luminous S0 galaxy with a large ring, UGC~5936, is studied here.}
   {By applying long-slit spectroscopy along the major axis of UGC~5936, we have measured gas and star
kinematics, Lick indices for the main body of the galaxy, and strong emission-line flux ratios in the ring. 
After inspecting the gas excitation in the ring using line ratios diagnostic diagrams
and having ensured that it is ionised mostly by young stars, we have determined the gas oxygen abundance by
using popular strong-line methods. Also we have proved the spatial proximity of the south-eastern
dwarf satellite to UGC~5936 and have measured its gas metallicity.}
   {The ionised gas of the ring is excited by young stars and has solar metallicity. Star formation in the ring
is rather prolonged, and its intensity corresponds to the current HI content of UGC~5936 (to the Kennicutt-Schmidt
relation). The whole morphology of the HI distribution implies current accretion of the cold gas from the satellite
onto the outer disc of UGC~5936; due to the satellite location and rotation in the plane of the stellar disc
of the host galaxy, the accretion is smooth and laminar providing the favourable condition for star formation ignition.}
   {}


   \maketitle
%

\section{Introduction}
Rings are common attributes of S0 galaxies by the definition of this morphological type. As it
was mentioned by \citet{devauc59}, ''Hubble distinguished two groups of S0 objects:
S0 (1), smooth lens and envelope;... S0 (2): some structure in the envelope in the form of a dark zone and ring...''
Surface photometry reveals that about 50\%\ S0-S0/a galaxies possess outer stellar
rings \citep{arrakis,nirs0s}. In the most cases the nature of these rings cannot be attributed to bar resonances,
oppositely to the way how it is commonly treated concerning a wider sample of spiral ringed galaxies \citep{buta86}. Just the occurence
of bars falls in S0s with respect to spiral galaxies: according to the statistics by \citet{lauri09}, only 46\%\ of S0s
show some signs of bars against 64\%--69\%\ among spirals, and these bars are mostly weak \citep{lauri10}.
Among the outer stellar rings, about 50\%\ are also seen in ultraviolet (UV) \citep{kostuk15} so betraying
recent star formation and being expected to contain probably some amount of gas to fuel
this star formation. The gas origin in S0s is still controversial; even less is known about star formation
in S0s providing stellar ring structures. In this Letter we will consider UGC~5936 -- a
northern-sky luminous (R)SA0$^+$ galaxy, $M_H=-24.86$ (NASA/IPAC Extragalactic database, NED). The optical-band ring and
the first spectral results for this galaxy were reported by \citet{kostuk75} and \citet{kostuk81} long ago.
We \citep{we_uvrings} noted its UV ring matching the geometry of the optical blue ring.
The galaxy was observed in 21~cm line and was found to be a rather gas-rich S0, with $1.38\times 10^9$\Ms\ of the
neutral hydrogen \citep{bluedisk1}. The distance to the galaxy is 108~Mpc, and the linear scale is
about 0.5~kpc per arcsec (NED). This paper is a part of
our series about the origin of starforming rings in S0 galaxies; earlier NGC~6534 and MCG~11-22-015 have been
described by \citet{s0ring1} and NGC~4513 by \citet{n4513_20}.

\section{Observations and the data involved}

Our long-slit spectral observations were made with a multi-mode focal
reducer SCORPIO-2 \citep{scorpio2} at the prime focus of the Russian 6-m
BTA telescope of the Special Astrophysical Observatory, Russian
Academy of Sciences. UGC~5936 was observed on February 28, 2014, with the $1''$-slit
aligned with the major axis, $PA(slit)=83\deg$, and the total exposure time was 75~min;
and also it was exposed in the orientation through the neighbouring dwarf galaxy,
$PA(slit)=135\deg$, with the exposure time of 45~min.
The seeing during these observations was mediocre, $FWHM=3$\arcsec.
We used the VPHG1200@540 grism providing an intermediate spectral
resolution FWHM $\approx 5$~\AA\ in the wavelength region from 3700~\AA\ to 7200~\AA.
This spectral range includes a set of strong absorption and emission lines
making it suitable to analyse both stellar and gaseous
kinematics of the galaxy as well as the gas excitation
and chemistry and the properties of the stellar populations. The slit is $6'$ in
length allowing to use the edge spectra to subtract the sky
background. The CCD E2V CCD42-90, with a format of $2048 \times 4600$,
using in the $1\times 2$ binning mode provided a spatial scale of 0.357\arcsec /px
and a spectral sampling of 0.86~\AA/px. The data reduction as well as kinematics 
and stellar population characteristics derivation were
standard for our SCORPIO-2 data -- see for example \citet{fp_s0} or \citet{n4513_20}.

To study the large-scale structure of the galaxy, we have used the $gr$-band images
from the SDSS/DR9 archive \citep{sdssdr9}.

\section{Photometric and dynamical structure of UGC 5936}

\begin{figure*}[htb!]
  \centering
\includegraphics[width=0.9\textwidth]{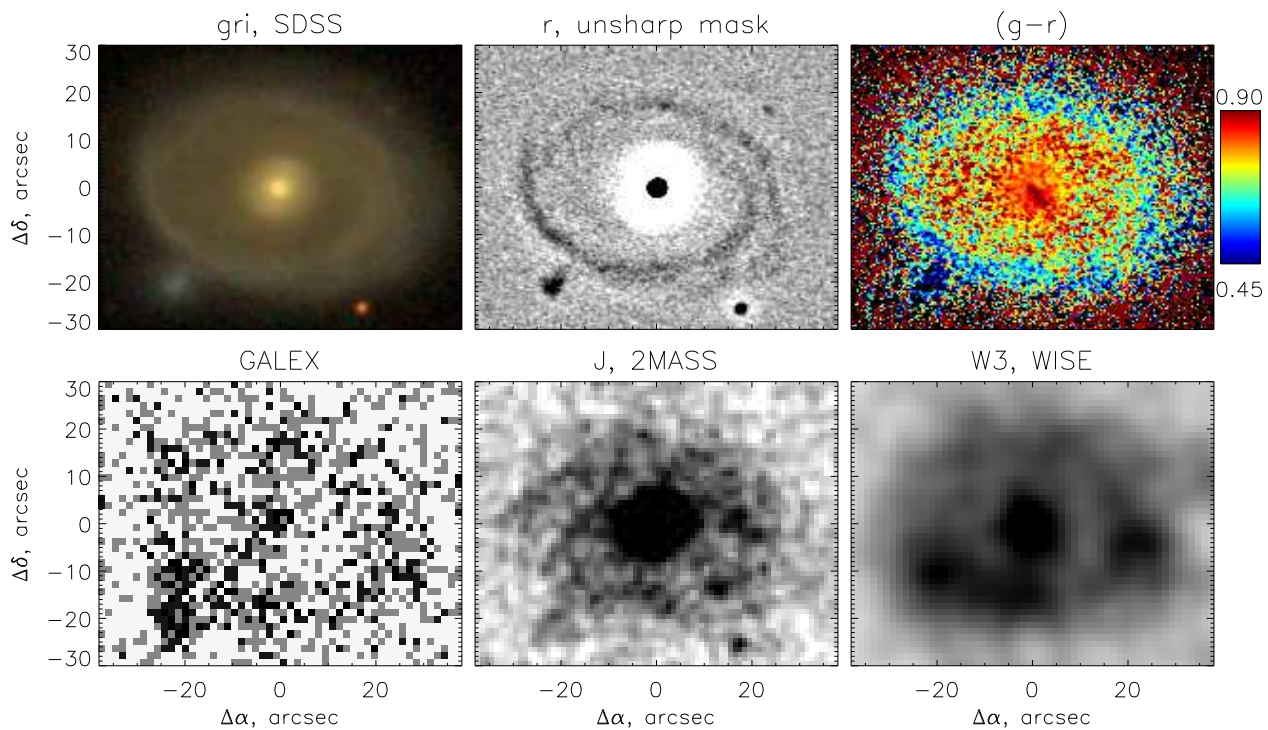}
\caption{All the maps available for UGC~5936. {\it The upper row}: the combined continuum-band optical image from
the SDSS, the r-band unsharp-masked image from the SDSS, and the (g-r) colour image. {\it The bottom row}: the GALEX NUV
image, the 2MASS J-band image, and the WISE/W3 12$\mu$m-band image.}
\label{images}
\end{figure*}

The smooth lenticular galaxy UGC~5936 (see 2MASS J-band image, Fig.~\ref{images}, {\it bottom middle}) possesses a large pseudoring
which looks unclosed in the optical continuum bands (Fig.~\ref{images}, {\it upper left and middle}).
However Fig.~\ref{images}, {\it bottom right},
which shows the WISE/W3 data in the 12$\mu$m band delineates a perfect closed ring; here hot dust heated by star formation
reveals its organisation in a typical S0 outer ring structure. The colour map (Fig.~\ref{images}, {\it upper right}),
demonstrates the blue colour of the ring, narrow to the east and splitting into two arcs to the west, as well as the red colour of
the arc connecting the core and the western part of the ring. The western blue-ring structures correspond to the ring in the
unsharp-masked image, Fig.~\ref{images}, {\it upper middle}. Interestingly, the GALEX NUV map (Fig.~\ref{images}, {\it bottom left})
which refers also to the young star concentration, shows UV flux excesses beyond the borders of the continuum-band pseudoring,
the offset is seen especially well at the major axis. To the south-east from UGC~5936 a small satellite,
SDSS J105009.10$+$362009.9, is observed in 15~kpc projected onto the sky plane (and in 19~kpc if we suggest that it lies
in the plane of the UGC~5936 disc); the satellite is very blue. According to SDSS/DR9, the magnitude difference between UGC~5936
and its satellite is $5.64^m$ in the $g$-band and $6.2^m$ in the $r$-band, so it is a bona-fide dwarf, with $M_B\approx -15^m$
and the stellar mass of about $10^8$~\Ms\ if we apply the mass-to-light ratio corresponding to its colour $g-r=0.27$ \citep{bell03}.

\begin{figure*}[htb!]
   \centering
   \centerline{
   \includegraphics[width=0.33\textwidth]{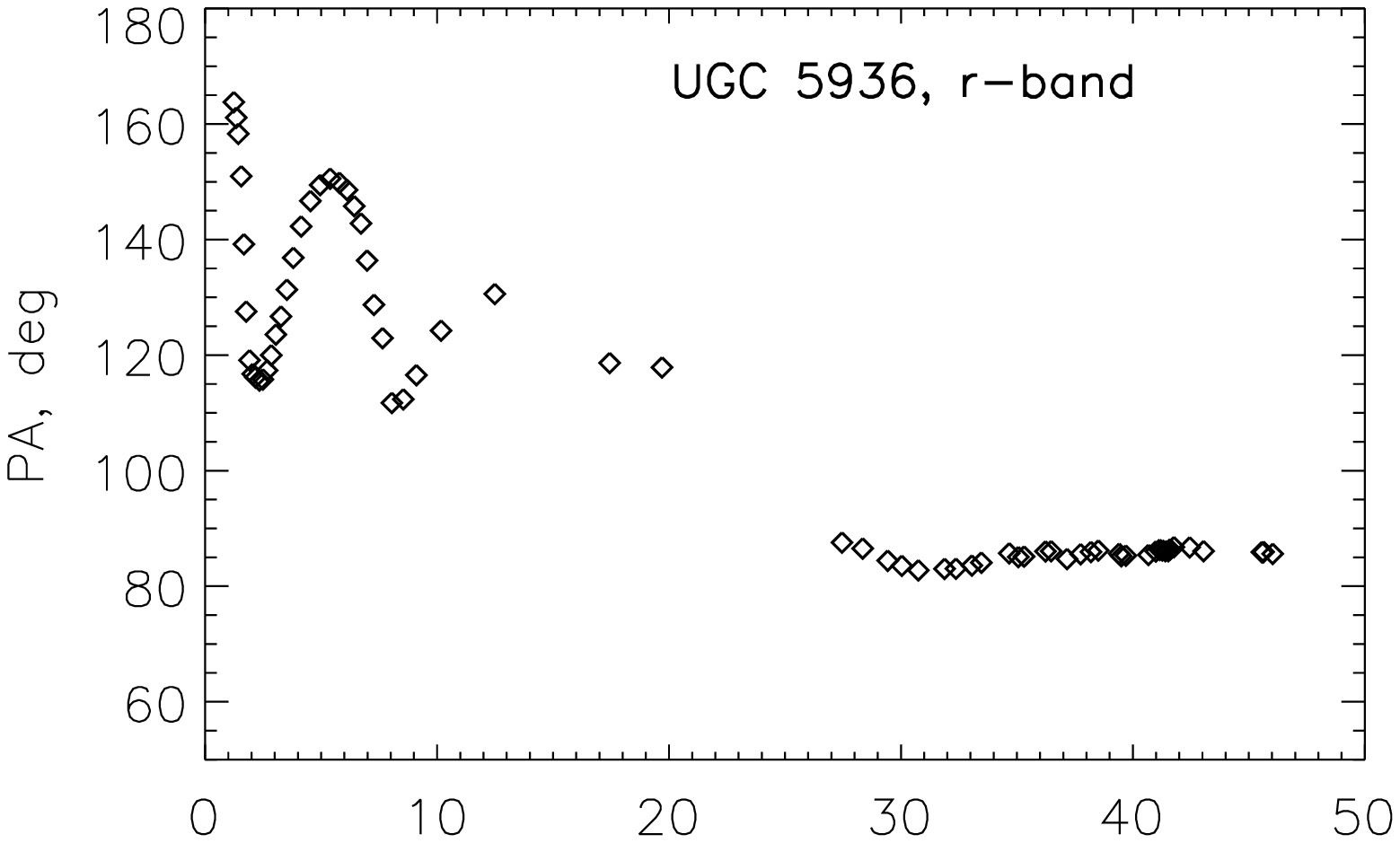}
   \includegraphics[width=0.33\textwidth]{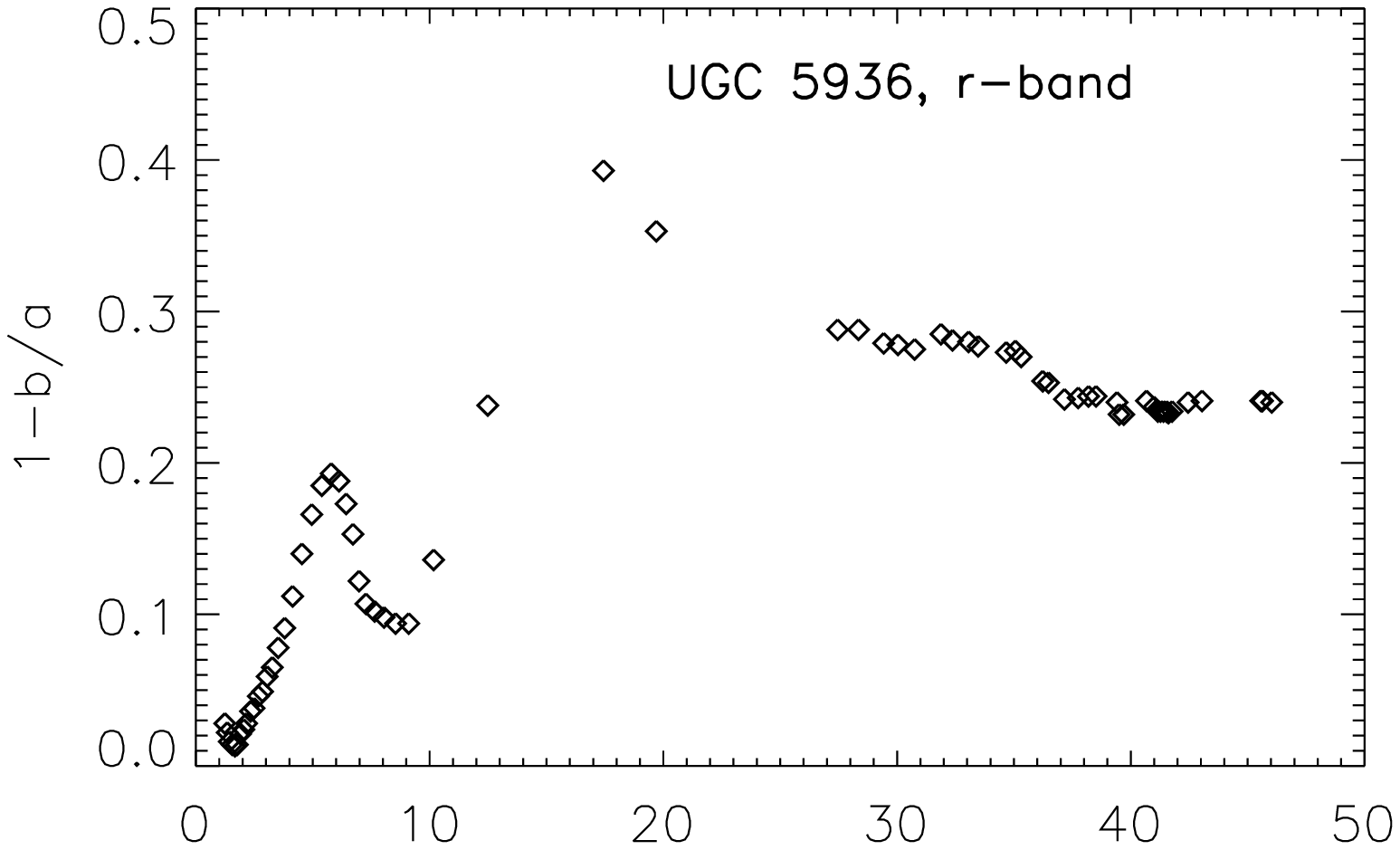}
   \includegraphics[width=0.33\textwidth]{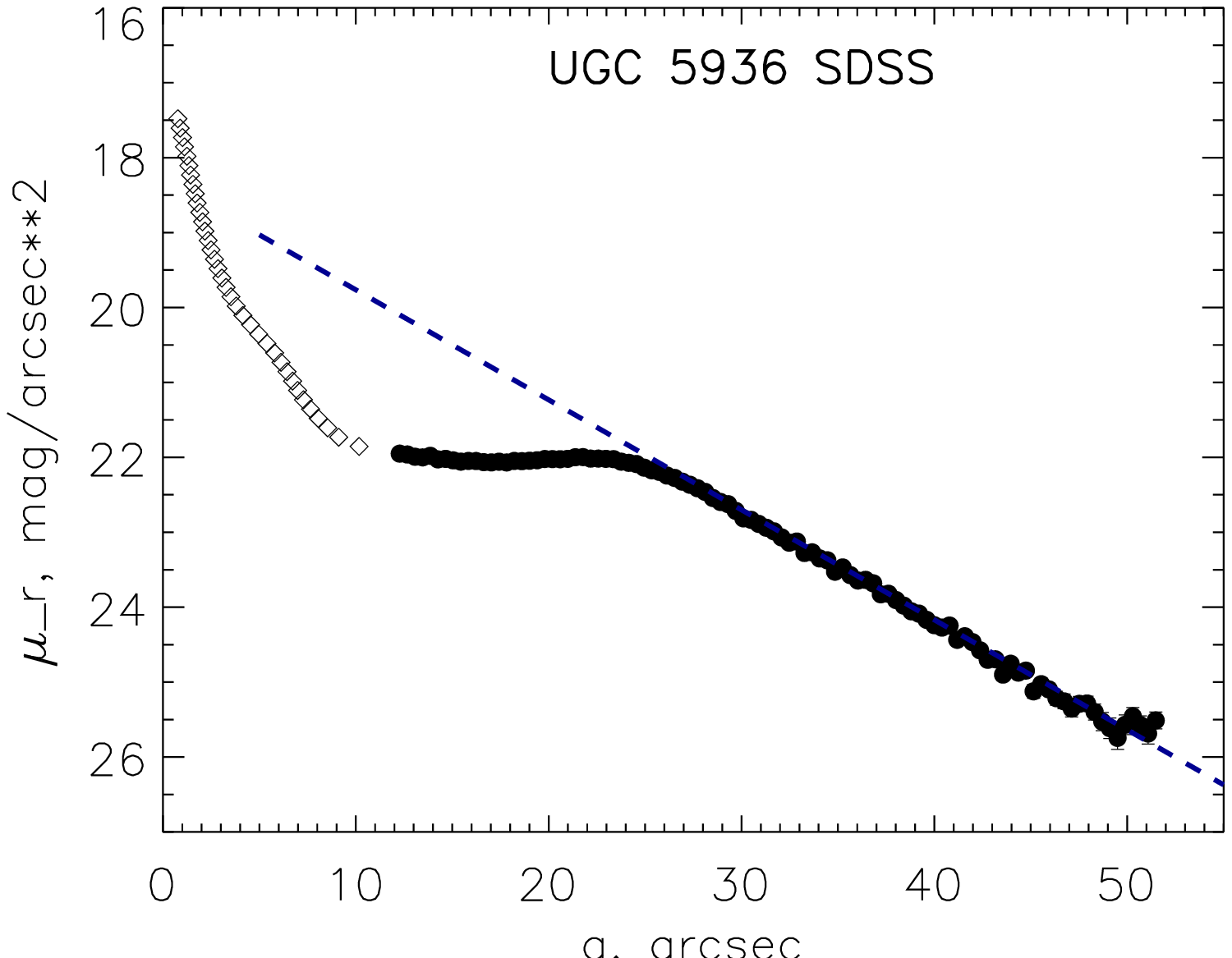}
   }
   \caption{The results of the photometric analysis of the SDSS/DR9 data. The blue dashed line
   in the right plot presenting the derived surface-brightness radial profile is a fitted
   exponential law showing the area of disc domination in UGC~5936.}
              \label{sdss5936}
\end{figure*}

    The ring of UGC~5936 could be classified as an outer ring since its radius is about 15~kpc. However the galaxy is very large,
and we trace certainly the isophote characteristics till about 22.5~kpc from the center. The results of the isophote analysis are
presented in Fig.~\ref{sdss5936}, two {\it left plots}, and in the {\it right plot} we show an azimuthally averaged radial surface brightness
profile in the $r$-band. At $R\approx 5\arcsec$ the local ellipticity maximum and major-axis position angle twist reveal a strong
triaxiality of the (pseudo)bulge. The regular behaviour of the isophote characteristics is observed at radii beyond $R=25\arcsec$
and indeed, starting from this radius outwards we see a classical exponential disc with a scalelength of $h=7.4\arcsec \pm 0.1\arcsec$,
or 3.5~kpc. However, it is a Freeman's Type~II profile \citep{freeman}: a large portion of the inner profile lies below the
continuation of the exponential fit inwards. In the radius range of 10\arcsec --20\arcsec\ (5--10~kpc), between the pseudobulge
and the exponential stellar disc, we see an extended `lens' component, with fairly flat surface brightness profile.

\begin{figure*}
\centering
\centerline{
\includegraphics[width=0.45\textwidth]{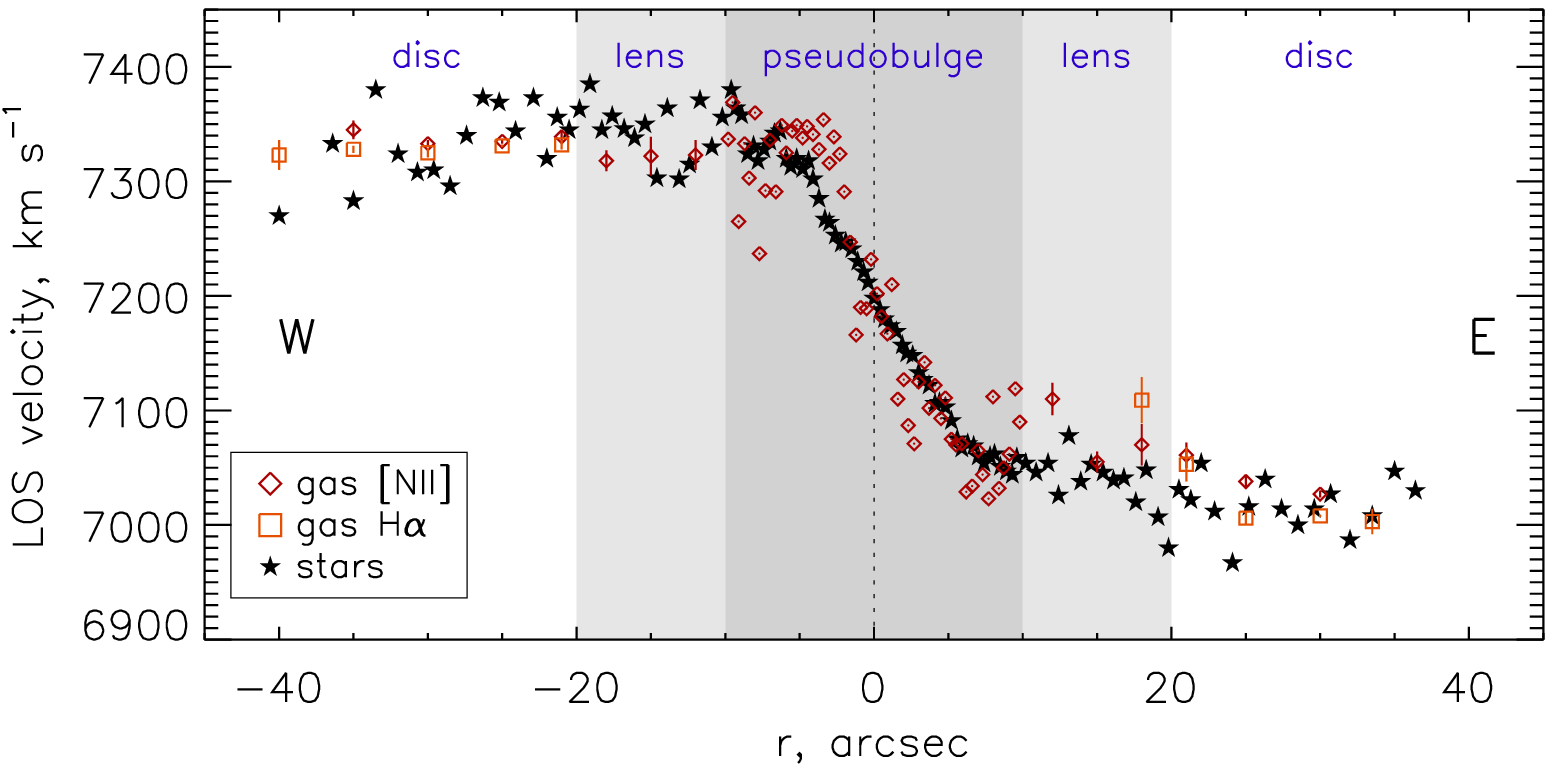}
\includegraphics[width=0.45\textwidth]{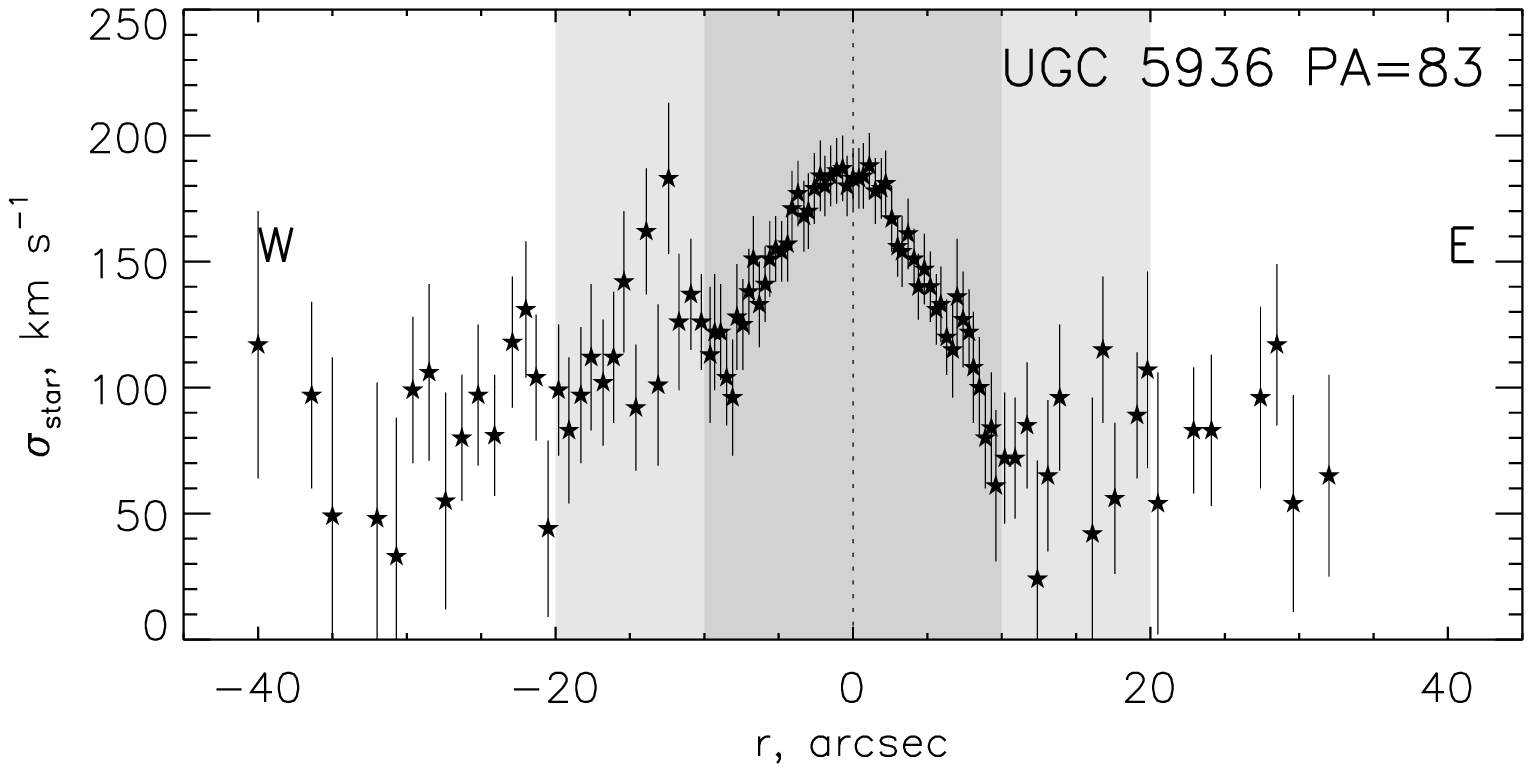}
}
\caption{The line-of-sight velocity profiles for the ionised gas and stars and the stellar velocity dispersion profile
in UGC~5936 along its major axis. In the left plot the stars show the stellar component while various red
signs relate to different emission lines of the ionised gas. The vertical shaded lanes designate different
structure components of the galaxy labeled in the left panel.}
\label{longslit_kin}
\end{figure*}

Interestingly, opposite to common expectations originating perhaps from classical works by John Kormendy \citep[e.g.]{n1553},
the lens of UGC~5936 is dynamically cold: by inspecting the profile of the stellar velocity dispersion (Fig.~\ref{longslit_kin},
{\it right plot}), we find that the stellar velocity dispersion in the lens is comparable to that in the outer stellar disc,
at least to the east from the nucleus. The western part of UGC~5936 is more perturbed by outer matter accretion as also
the gas component demonstrates -- see below the Section~5 about ionised-gas excitation.
The rotation velocity profiles of the gas and stars (Fig.~\ref{longslit_kin}, {\it left plot}) are identical so we may be sure
that the ionised gas rotates in the main galactic plane sharing the spin with the stellar disc. The line-of-sight velocity
of the satellite, 7045~\kms, is very close to the outer velocities of the stars and gas in the disc of UGC~5936, 7100~\kms\ and
7080~\kms\ respectively, so the orbital spin of the satellite is probably similar to the rotation spin of UGC~5936. It is
a very favourable configuration for smooth gas accretion from the satellite into the disc of UGC~5936.

\section{Stellar population properties}

\begin{figure*}[htb!]
   \centering
   \centerline{
   \includegraphics[width=0.33\textwidth]{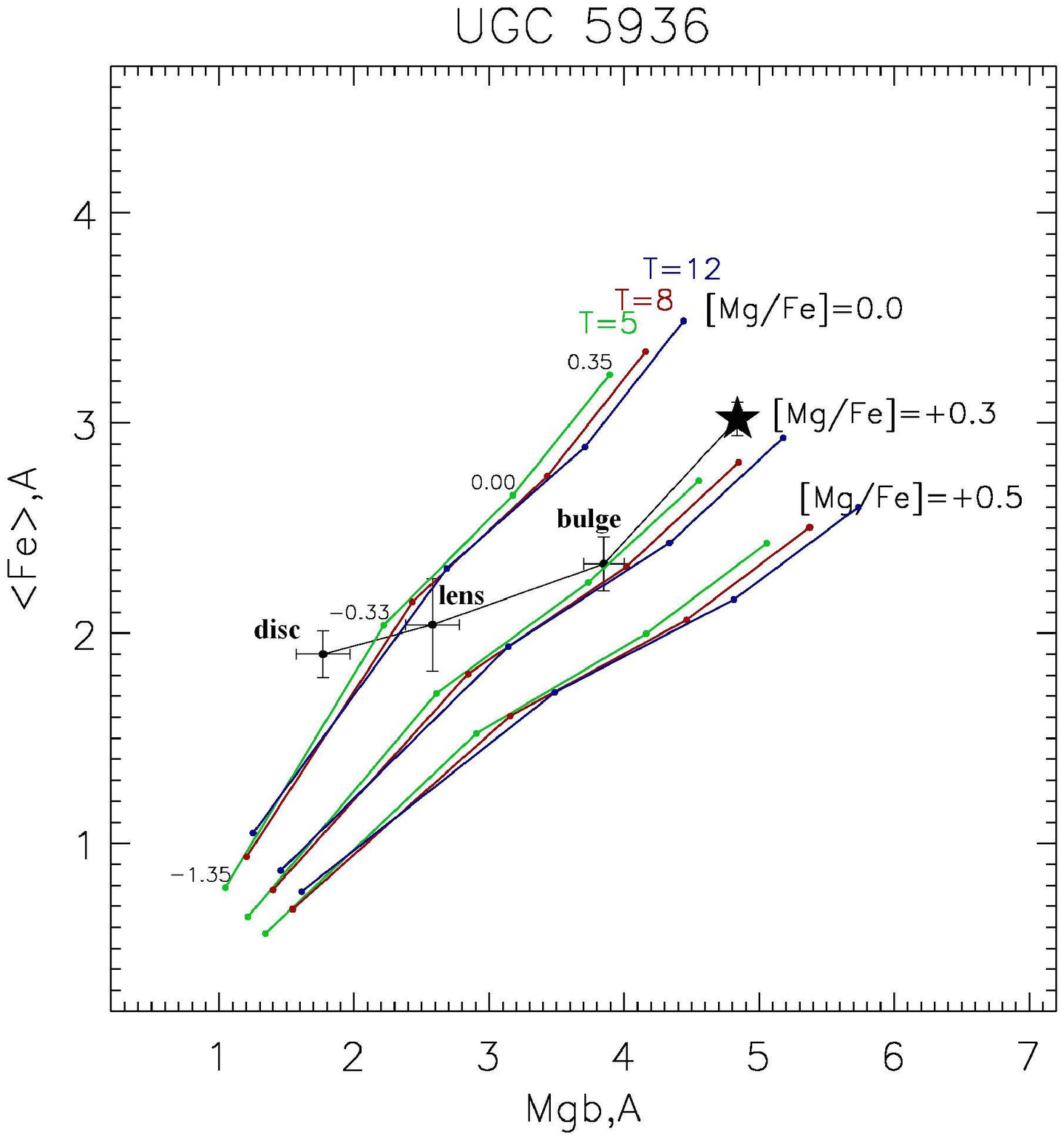}
   \includegraphics[width=0.33\textwidth]{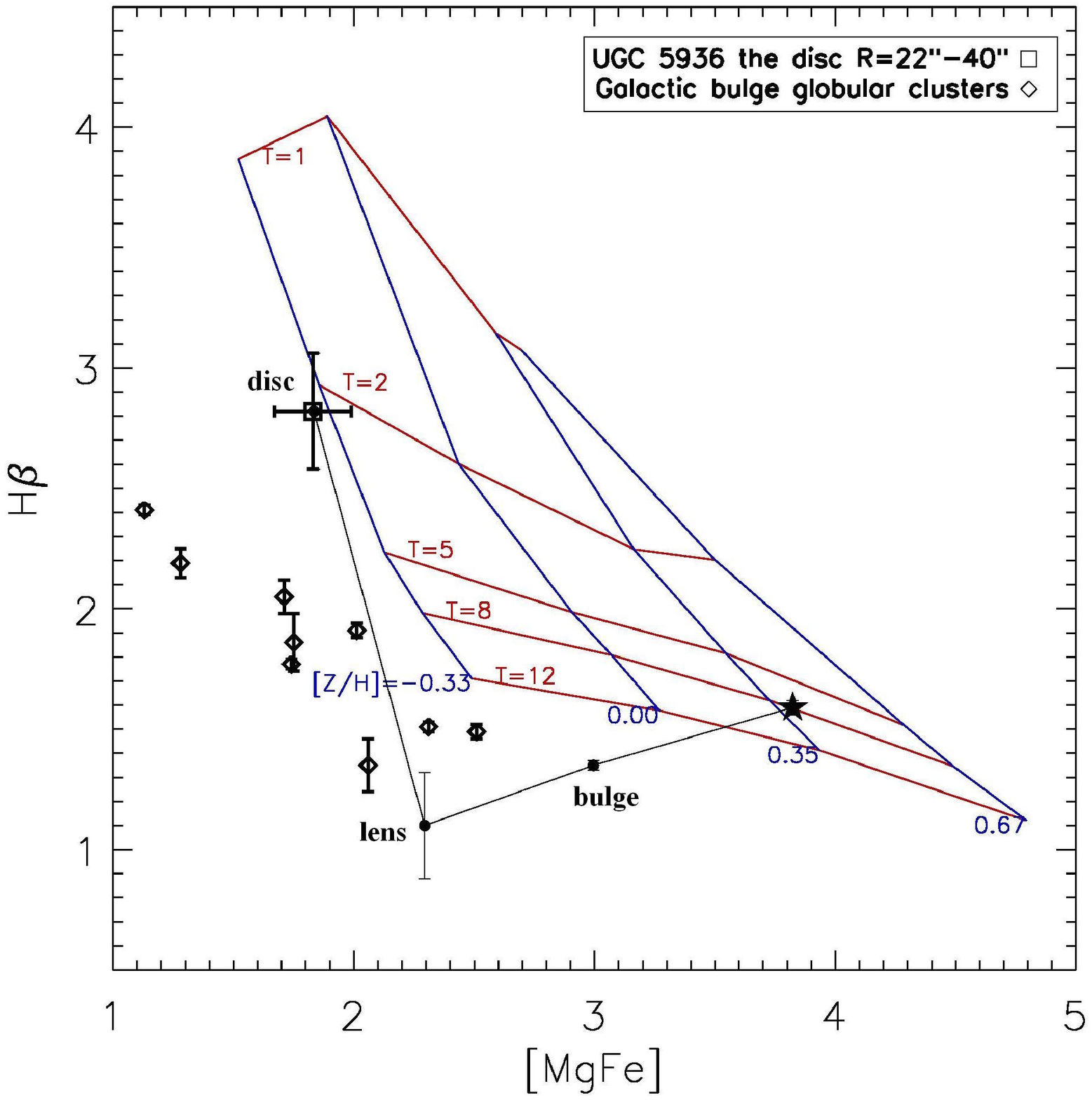}
   }
   \caption{Lick index-index diagrams for UGC~5936. 
The {\it left plot} represents Mgb vs iron index diagram which allows to estimate magnesium-to-iron ratio through the comparison of
our measurements with the models by \citet{thomod} for the different Mg/Fe ratios. By confronting the H$\beta$ Lick index versus
a combined metallicity Lick index involving magnesium and iron lines ({\it right plot}), we solve the metallicity-age degeneracy
and determine these stellar-population parameters. Five different age sequences (red lines) are plotted as reference frame; 
the blue lines crossing the model age sequences mark the metallicities of $+0.67$, $+0.35$, 0.00, --0.33 from right to left.
A large black star corresponds to the central core, and then we go along the radius through the galaxy structure components:
$R=4^{\prime \prime}-10$\arcsec (pseudobulge), $R=10^{\prime \prime}-20$\arcsec (lens), and $R=20^{\prime \prime}-40$\arcsec (disc).
A few globular clusters from \citet{kim_glcl} belonging to the Galactic bulge and thick disc
are also plotted to delineate the empirical old-age sequence for the metallicities below [Z/H]$=-0.4$.}
\label{lickind}
    \end{figure*}

We have estimated the properties of the stellar population along the radius of the major-axis cross-section of UGC~5936
by using Lick indices. The results are presented in Fig.~\ref{lickind}. The magnesium-to-iron ratio characterising duration
of the star formation \citep{matt86,matt94} is supersolar in the nucleus and in the bulge, but it is solar in the lens and
in the disc giving again the evidence in favour of disc origin of the lens. And the stellar metallicities of the lens and
of the disc are similar, about one half or one third of the solar. But the mean stellar ages are strongly different:
the lens is very old, older than 10~Gyr, comparable to the Galactic globular clusters, and the mean age of the stellar population
in the inner disc (where the contribution of the blue ring is significant) is only about 2~Gyr.

\section{Gas-phase metallicity}

\begin{figure*}
\centering
\centerline{
\includegraphics[width=0.33\textwidth]{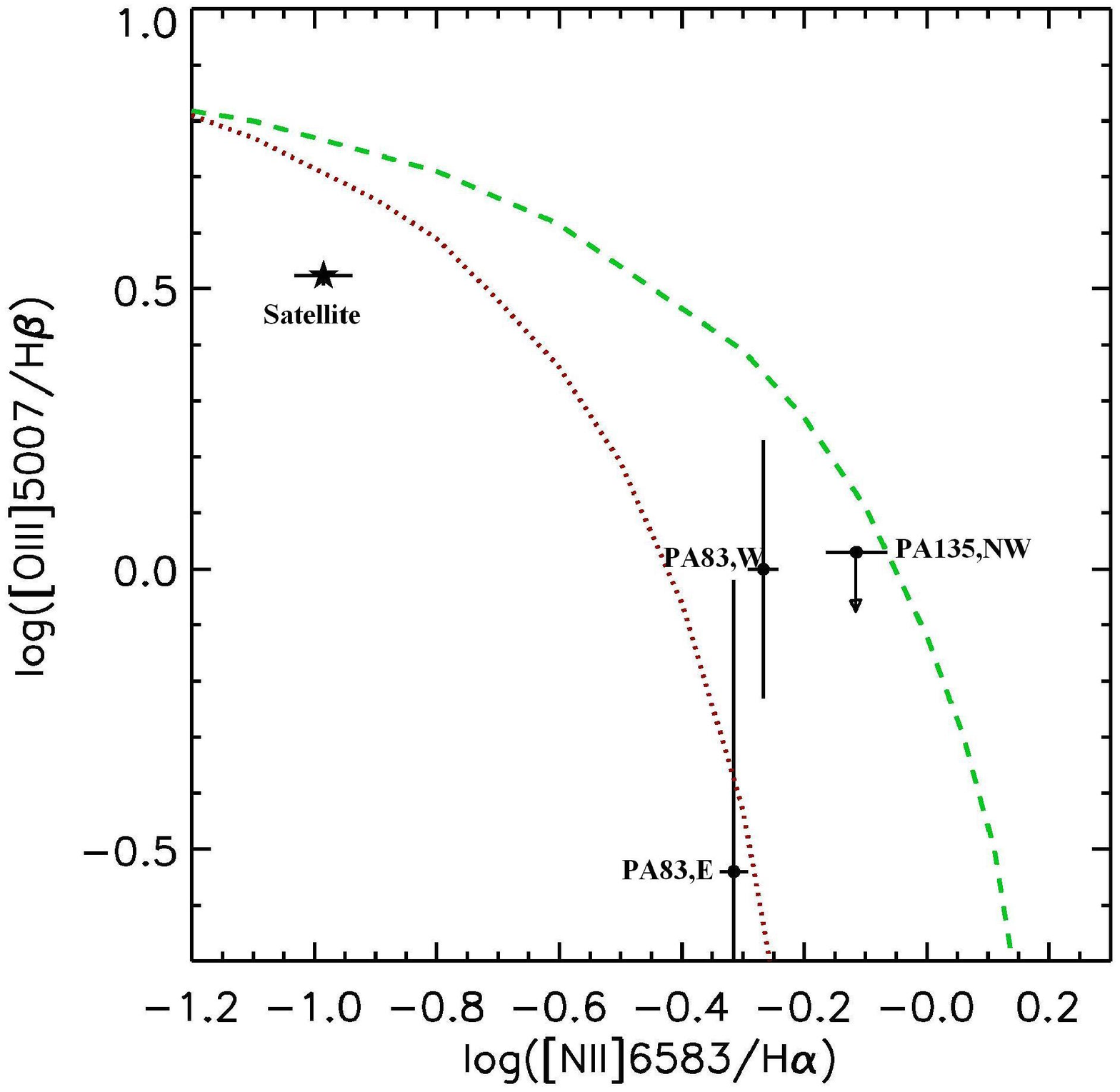}
\includegraphics[width=0.33\textwidth]{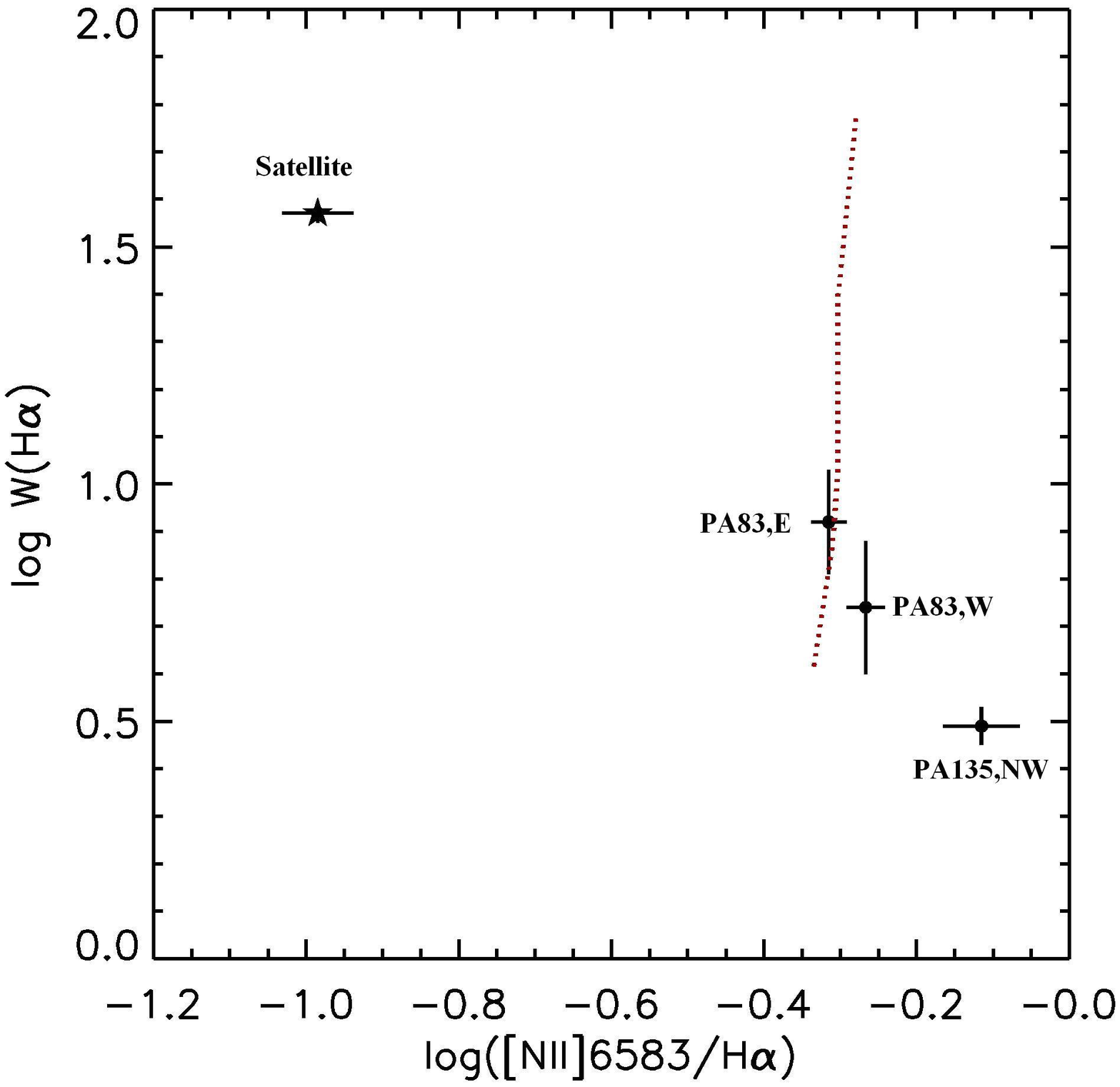}
}
\caption{Diagnostic diagrams for gas excitation determination: the classical \citet{bpt}' diagram in the {\it left plot} and an advanced
version proposed by \citet{cid2010} including the H$\alpha$ emission-line equivalent width in the {\it right plot}. The bordering lines
between the young-star excitation, to the left of them, and the other mechanisms are those from \citet{kewley01} (green dashed line)
and from \citet{kauffmann03} (red dotted line). The black points represent the various parts of the UGC~5936 ring; the black star
is the measurement of the south-eastern dwarf satellite.}
\label{bpt}
\end{figure*}

In the main body of UGC~5936 the emission line \Ha\ is not detected, and the strongest emission line is [NII]$\lambda$6583.
Such emission-line pattern is consistent with the possible gas excitation by old stars \citep{pagb94,liermodel} that is in agreement
with the age of the stellar population in the central part of UGC~5936. In the ring where strong H$\alpha$ emission lines are seen
at the radii $R=28\arcsec$ to the east and $R=25\arcsec$ to the west, we expect the gas excitation by current
star formation. We have plotted the emission-line ratios for the ring onto the Baldwin-Phillips-Terlevich (BPT) diagram (Fig.~\ref{bpt},
{\it left}). Indeed, the strong-line ratios in the ring of UGC~5936 have appeared to lie below the theoretical border
of the star formation calculated by \citet{kewley01} so the ionised gas of the ring may be mostly excited by young stars.
But the offset of the western part of the ring at the BPT-diagram with respect to the observational star formation sequence
by \citet{kauffmann03} puts this region into the so called 'composite zone' revealing a possible contribution of shocks 
into the excitation of the ionised gas of the ring to the west from the nucleus. The other variant of the diagnostic
diagram including the \Ha\ equivalent width (Fig.~\ref{bpt}, {\it right}) shows the same separation: only the eastern
side of the ring is fairly consistent with the \citet{kauffmann03}'s star formation sequence in the \citet{cid2010}'s modification.
The outermost arm of the western part of the ring, at $R=28\arcsec$ to the west from the nucleus, is certainly excited
by shock because its $\log (\mbox{[NII]}\lambda 6583 /\mbox{H} \alpha ) =-0.09 \pm 0.02$.

We have also plotted in Fig.~\ref{bpt} the measurements of the ionised-gas emission-line intensity ratios for the dwarf
south-eastern satellite which has been probed by our long-slit cross-section in $PA=135^{\circ}$. Here the H$\alpha$
emission line dominates in the spectrum, with the equivalent width of 38~\AA, and the gas is unanimously
excited by young stars. By taking into account this fact, we have determined the gas oxygen abundance by using the strong-line method.
By exploring both O3N2 and N2 calibrations from \citet{pp04} and \citet{marino13}, we have obtained for the satellite 
$12+\log \mbox{(O/H)} =8.32\pm 0.05$~dex. We must note that this value is higher by about 0.2~dex
than the mean metallicity implied by the mass-metallicity relation for the galaxy of $M_*=10^8$~\Ms\
as it is calibrated by \citet{kewley_ellison} through the \citet{pp04}' approach.

For the ring of UGC~5936, precisely, for its eastern part where there is no doubts on the HII-like excitation,
the same calibrations of the strong-line metallicity indicators give $12+\log \mbox{(O/H)} =8.64\pm 0.05$~dex,
or nearly solar oxygen abundance.

\section{Discussion and conclusion: Star formation and origin of the ring in UGC 5936}

As we have mentioned in Section 3, UGC~5936 is well seen in the 12$\mu$m-band of the WISE Allsky survey. This fact has
allowed \citet{chang_sf} to determine its star formation rate (SFR) by analysing the full SDSS$+$WISE spectral energy
distribution (SED): the integrated SFR of UGC~5936 is equal
to 0.24~\SFR. With the total stellar mass of $10^{11}$~\Ms\ \citep{chang_sf}, it is well below the main sequence.
However when we inspect the data on the neutral hydrogen in UGC~5936, with the estimate of $\Sigma (HI) =1$~\Ms\ per pc$^2$
\citep{bluedisk1}, and compare it with the SFR surface density (by dividing the total SFR by the galaxy area within
$R_{25}$), we ensure that the galaxy is exactly on the Kennicutt-Schmidt relation \citep{ken98,chang_sf}. The outer
disc of the galaxy experiences quite normal star formation.

The map of the neutral hydrogen distribution in UGC~5936 presented by \citet{bluedisk1} is very curious: it demonstrates
two condensations -- one centered onto the western part of the UGC~5936 ring and another centered onto the south-eastern satellite.
The latter condensation is partly projected onto the eastern part of the UGC~5936 ring. The whole view is consistent with the
HI ring in the disc plane of UGC~5936, and the probable source of the gas is the satellite. We have found that the metallicity of
the satellite is higher than expected for its present stellar mass; it is perhaps because the satellite is already mostly
disrupted by UGC~5936 having accreted both the gas and the stellar component of the satellite. The accretion proceeds in a laminar
regime because the satellite co-rotates the outer disc of UGC~5936, and it results in quite normal,
for its gas content, star formation rate within the accreted gaseous ring. \citet{vdvoort} reported
an opposite effect, namely, the galaxy locations below the Kennicutt-Schmidt relation found for six S0s experiencing
recent gas-rich minor merging. However, they investigated S0s with disturbed morphology, and the initial orbital configuration
of the satellite merging remained then unknown. Perhaps, their particular sample galaxies experienced off-plane satellite accretion;
in these cases star formation indeed had to be suppressed \citep{fp_s0}.

The accreted gas has made at least two revolutions around UGC~5936: two starforming arcs with different radii are seen
at the western part of the ring, and the gas excitation there includes some contribution from shock waves implying
collision of gas flows. The star formation in UGC~5936 -- and the satellite disruption -- has then begun 1--2 Gyr ago
since the orbital time at the radius of the ring is about 0.5~Gyr. Such prolonged star formation event can also explain the solar
metallicity of the ionised gas in the ring of UGC~5936 though the initial oxygen abundance of the accreted gas is lower
by a factor of 3: evidently, the chemical evolution stage of the ring is very advanced resulting in the gas metallicity saturation
\citep{ascasibar15,zahid14}.

So UGC~5936 represents a certain example of the outer S0 ring which origin is outer gas accretion. The source of the cold gas inflow can
be firmly established in this particular case: it is a dwarf gas-rich satellite which may perhaps be completely disrupted after
a few orbital periods. It may be a common way to form outer starforming rings in S0 galaxies without bars. We have visually inspected
close environments of the UV-detected outer rings in S0s from the compilation by \citet{kostuk15}. Within 50~kpc from a host,
70\% $\pm$10\%\ unbarred S0s with UV-rings have a blue dwarf satellite while among barred S0s with UV-rings -- only 27\% $\pm$ 14\%\
(the errors are estimated from the binomial distribution). So the outer rings in barred S0s may be formed due to gas accumulation at the
outer Lindblad resonances as it has been argued by \citet{atha82} and \citet{buta86}; but for unbarred S0s the outer ring origin
through gas-rich dwarf merging seems to be more probable.

\begin{acknowledgements}
We thank the anonymous referee who has made very useful comments resulting in the paper improvement.
We are grateful to Roman Uklein for the help during the observations.
The study of galactic rings was supported by the Russian Foundation for Basic
Researches, grant no. 18-02-00094a. The work is based on the data obtained at the Russian
6m telescope of the Special Astrophysical Observatory operated with the financial support of the Ministry of Science
and Higher Education of the Russian Federation (including agreement No. 05.619.21.0016, project ID RFMEFI61919X0016)
and on the public data of the SDSS (http://www.sdss3.org) survey.
This research has made use of the NASA/IPAC Extragalactic Database (NED), which is funded by the National Aeronautics
and Space Administration and operated by the California Institute of Technology.
The NASA GALEX mission data were taken from the Mikulski Archive for Space Telescopes (MAST).
The WISE data exploited by us were retrieved from the NASA/ IPAC Infrared Science Archive,
which is operated by the Jet Propulsion Laboratory, California Institute of Technology,
under contract with the National Aeronautics and Space Administration.
\end{acknowledgements}

%
%

\end{document}